\documentclass[a4paper,11pt]{article}

\usepackage{pos}
\usepackage{lipsum} 
\usepackage{float}
\usepackage{subcaption}
\usepackage{lineno}

\title{A Search for Astrophysical Neutrinos from Flaring X-ray Binaries with IceCube}

\ShortTitle{Astrophysical Neutrinos from Flaring X-ray Binaries}

\author{The IceCube Collaboration \\{\normalsize \normalfont(a complete list of authors can be found at the end of the proceedings)}\\}

\emailAdd{kochocki@msu.edu}
\emailAdd{doddamre@msu.edu}
\emailAdd{nisamehr@msu.edu}
\emailAdd{ssclafani@icecube.wisc.edu}

\abstract{
Recently, IceCube has observed an excess of astrophysical neutrinos from the Galactic plane. Such a signal may indicate the presence of individual sources or a diffuse neutrino flux from the interactions of Galactic, hadronic cosmic rays. We consider the prospect of neutrino production in galactic X-ray binary systems. A model for neutrino production in the variable jets of these systems is discussed, and how such information may be incorporated within searches for astrophysical neutrinos from these sources. Two ongoing studies are presented on behalf of the IceCube Collaboration using 13 years of cascades and track-like events. In the first study, recently published model predictions are used to motivate the analysis of five selected black-hole X-ray binaries. Gamma-ray data from Fermi-LAT, as well as X-ray data from Swift/BAT and MAXI are used to search for temporal correlations between radiative cycles of the disk-jet system and potential neutrino emission. In the second study, a wider selection of variable X-ray binaries and their Swift/BAT light curves is considered and a stacking search is used to constrain the contribution of this source population to the Galactic neutrino flux.

\vspace{4mm}

{\bfseries Corresponding authors:}

Alina Kochocki$^{1*}$, 
Dhiti Doddamreddy$^{1}$, 
Mehr Un Nisa$^{1}$\\
Steve Sclafani$^{2}$\\

{$^{1}$ \itshape Michigan State University}\\
$^{2}$ \itshape University of Maryland\\[4mm]
$^*$ Presenter
}

\ConferenceLogo{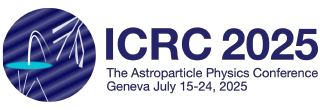}

\FullConference{39th International Cosmic Ray Conference (ICRC2025)\\
 15–24 July 2025\\
Geneva, Switzerland\\}

\begin{document}

\maketitle
\section{Introduction}\label{sec1}
The IceCube Neutrino Observatory, situated at the geographic South Pole, has established an approximately isotropic diffuse flux of high-energy astrophysical neutrinos \citep{Aartsen_2013a}. In recent years, a component of this flux has been observed from the galactic plane \citep{gp_2023}. Such neutrinos may originate from the interactions of galactic cosmic rays with interstellar gas. However, individual, unresolved astrophysical sources can also contribute to this neutrino excess from the galaxy. 

Galactic X-ray binaries (XRBs) are promising candidate sources of neutrinos and are of particular interest for transient neutrino searches. XRBs are composed of a compact object—typically a stellar-mass black hole or neutron star—and a stellar companion \citep{Remillard_2006}. Material of the companion is accreted onto the compact object, forming an accretion disk and in some cases powering the ejection of relativistic jets. These systems are relatively close to Earth, have well-measured distances, and show significant variability across the electromagnetic spectrum, especially in X-rays and gamma rays. This combination of high-energy activity and proximity makes them strong candidates for targeted neutrino observations.

\section{X-ray Binary Hysteresis}\label{sec3}

An understanding of the cyclic nature of X-ray binary accretion and jet production has developed over the last few decades. Initially recognized as X-ray bright sources, early studies focused on interpretation of their varying X-ray spectra \citep{Remillard_2006}. Such sources are well characterized by two emitting regions – a variable accretion disk with thermal and non-thermal radiation, and a radio and gamma-ray bright jet. X-ray activity generally traces different states of the accretion disk. It is somewhat unclear if observed non-thermal emission relates to scattered photons around the disk, from a forming jet, or if non-thermal coronae around the disk is synonymous with a jet base \citep{xrb_jet_article}.

X-ray binaries exhibit distinct X-ray spectral states. At first, luminous, thermal emission from the disk may be observed. Softer, lower-energy X-ray emission exceeds harder, higher-energy X-rays in intensity. Later, a dim, soft power-law spectrum of non-thermal X-ray emission may develop. This non-thermal component continues to grow, transitioning to a hardened power-law spectrum. Jets are commonly observed at this time. Eventually, the non-thermal emission may shift to a bright, soft power-law spectrum, where sporadic jets could still be observed. The source then redevelops a strong thermal emission component, repeating this cycle.   

The four states described above trace a single cycle of X-ray binary accretion and jet production. These states have been confirmed for a variety of X-ray binaries, though exceptions to this cycle have been observed. In general, emission should vary on this cyclic timescale. This is in addition to any modulation due to the periodic nature of the binary system. 

\section{Neutrino Production in X-ray Binaries}\label{sec4}

Recent work has modeled neutrino production from the jets of selected X-ray binaries \citep{Kantzas_2023}. A direct link between the hot, thick accretion disk with non-thermal emission and the jet has been proposed \citep{xrb_jet_article}. Photons of the original thermal accretion disk may be up-scattered on a coronal plasma in the inner regions. The accelerated electrons and protons forming this corona could represent the spectrum of material eventually entrained within a jet (i.e., the corona and jet-base material are synonymous). Multiwavelength data for two sources has been treated – the high-mass X-ray binary, Cygnus X-1, and low-mass X-ray binary, GX~339-4. 

With accelerated protons in the jet, hadronic interactions via p-p and p-$\gamma$ channels may proceed, supported by thermal and non-thermal photon fields, and protons from stellar winds of the companion object. Such neutrino emission could correlate with jet formation, gamma-rays from leptonic or hadronic interactions in the jet, and the observed X-ray hard states. 

More recently, neutrino emission has been associated with Seyfert active galactic nuclei \citep{ngc_1068, x_ray_bright_seyferts}. The X-ray bright corona around the accretion disk could provide the high-energy photons necessary for efficient p-$\gamma$ neutrino production \citep{Kheirandish_2021}. Such a model may also extend to X-ray binaries. In this way, high-energy neutrinos may come from both the region of the accretion disk as well as the jet.

\section{Selected X-ray Binaries}\label{sec5}

We consider predictions of astrophysical neutrino intensity from the jets of galactic black hole X-ray binaries. A leptohadronic X-ray binary model has been applied to both Cygnus X-1 and GX~339-4 in Ref. \citep{Kantzas_2023}, as representative high and low-mass X-ray binaries, respectively. These predictions represent the rate of neutrino production during jet-active states. Using scaling relations related to object distance, black hole size and jet power, predictions of expected neutrino rates were extended to 35 known galactic X-ray binaries \citep{Kantzas_2023}. 

The interacting proton luminosity may have a few order-of-magnitude uncertainty within the accelerated jet, from stellar winds and potential contributions from the disk. We assume a jet-duty cycle or jet active state of of 10$\%$ and compare to IceCube sensitivities for simulated sources of comparable variability and location. Details of the IceCube data analysis are provided in the following sections. Five sources are found with sensitivity fluxes for a power-law neutrino spectrum of index 2 within four orders of magnitude of the source flux prediction. These sources are described in Table \ref{tab:info}. We consider these sources as interesting candidates and targets for individual analyses. 

In this analysis, we utilize the time variability of each selected source in a search for astrophysical neutrinos. Soft and hard X-ray light curves from the MAXI ($\sim$2--20 keV) and Swift ($\sim$15--50 keV) satellites are used, respectively \citep{Mihara_2022, Krimm_2013}. Additionally, soft and hard gamma-ray light curves from Fermi-LAT are utilized (approximately 0.1--0.8 GeV and 0.8--10 GeV, respectively) \citep{Abdollahi_2023}. 

\begin{table}[]
\centering
\label{tab:source_table}
\begin{tabular}{ccccc} 
\hline 
Source Name & R.A., Dec. [\textdegree] & Period & $M_\odot$ & Long-Term Monitoring \\
\hline 
Cygnus X-3 \cite{cygx3_mass,cygx3_period} & 308.11, 40.95 & 4.8 hours & 8-14  & Fermi, MAXI, Swift \\
Cygnus X-1 \cite{cygx1_mass,cygx1_period} & 299.59, 35.20 & 5.6 days & 21 ± 2.0  & Fermi, MAXI, Swift \\
V4641 Sagittarii \cite{v4641_sgr_period_mass} & 274.84, -25.43 & 2.8 days & 6.4 ± 0.6  & Fermi, MAXI, Swift \\
MWC 656 \cite{mwc656_period_mass} & 340.73, 44.72 & 60.4 days & 3.8-6.9  & Fermi, MAXI \\
MAXI J1836-194 \cite{maxi_j1836_194} & 278.93, -19.32 & $<$ 4.9 days & 7.5-11  & Fermi, MAXI, Swift \\
\hline 
\end{tabular}
\caption{Properties of the five sources selected for analysis based on their predicted neutrino emission. We provide the common name, right ascension and declination (R.A. and Dec.) in degrees, approximate system period, compact object mass in solar masses ($M_\odot$) and monitoring coverage based on instrument for each source. Corresponding references are listed within the table. }
\label{tab:info}
\end{table}

\section{Operation and Data Selection}\label{sec6}
The IceCube Neutrino Observatory is a cubic kilometer of instrumented ice at the geographic South Pole. An array of 5,160 photosensors detects the radiative signatures of particle interactions within the ice. Data trigger and readout has been discussed in previous works \citep{icecube_technical}. 

IceCube observes the interactions of astrophysical neutrinos. Neutral-current interactions as well as electrons and positrons from charged-current interactions deposit observable radiation within the ice. These near-spherical radiative signatures, or cascades, are associated with poor angular resolution and excellent energy resolution. Muons or anti-muons which can be produced in charged-current interactions decay slowly and may deposit radiation along a track for multiple kilometers. Muons are associated with excellent angular resolution and poorer energy resolution. 

Atmospheric muon bundles and neutrinos are also produced locally in cosmic-ray air showers. Atmospheric muons from the north are largely diminished due to interaction within the earth. 

In this analysis, we utilize a combined set of selected track and cascade-like events. These data selections are largely constructed on the basis of event topology, partially reducing the contribution of atmospheric background. Details of these selections have been previously published \citep{point_source_selection, gp_2023}. 

\section{Analysis}\label{sec7}
To investigate potential neutrino emission from the target X-ray binaries, we conduct a time-dependent likelihood analysis using a number of temporal models specifically crafted for each source. The approach incorporates general knowledge about the emission patterns of each system in a search for astrophysical neutrinos correlated with with gamma-ray and X-ray light curves.

Gamma-ray emission is expected to trace the interactions of accelerated, high-energy particles within the jet. Fermi-LAT gamma-ray observations are used to construct temporal models for a correlation with neutrino data. Where available, we consider both hard and soft bands of gamma-ray data, potentially probing emission from areas of varying opacity. As gamma-rays can originate from partially opaque regions of the source, we assume that neutrinos could arrive slightly before or after the observed gamma-ray peaks. We allow for a relative lag between X-ray data and neutrino data as a free parameter in fitting. The lag is limited to about $\pm 50$ days, comparable to the shortest timescale of flaring variability observed from these sources. 

Traditionally, hard X-ray states or transitions from the hard X-ray state (increased hard X-ray intensity) are expected to trace jet formation. However, exceptions to this pattern exist. In the case of Cygnus X-3, gamma-ray and radio-bright jets are instead observed during soft X-ray maxima and transitions to this state. As it is expected that XRB emission is modulated on the timescale of accretion disk hysteresis, we assume a relative lag between X-ray data and neutrino data on the timescale of the observed hysteresis cycle. Both hard X-ray data from Swift and soft X-ray data from MAXI is utilized. 

With each light curve of our selected sources, we average individual observations within equally-spaced time bins. As all of our data comes from satellite monitoring programs, the cadence is fairly regular over the relevant IceCube observing period. The bin size is adjusted for each source based on the apparent characteristic scale of variability and statistical uncertainty. We use this resulting temporal probability density function (PDF) in our correlation with IceCube data. We find that this individualized method gives comparable or superior control in PDF construction compared to a Bayesian approach for a small set of light curves. 

\begin{figure}[]
\centering
\includegraphics[width=0.95\linewidth]{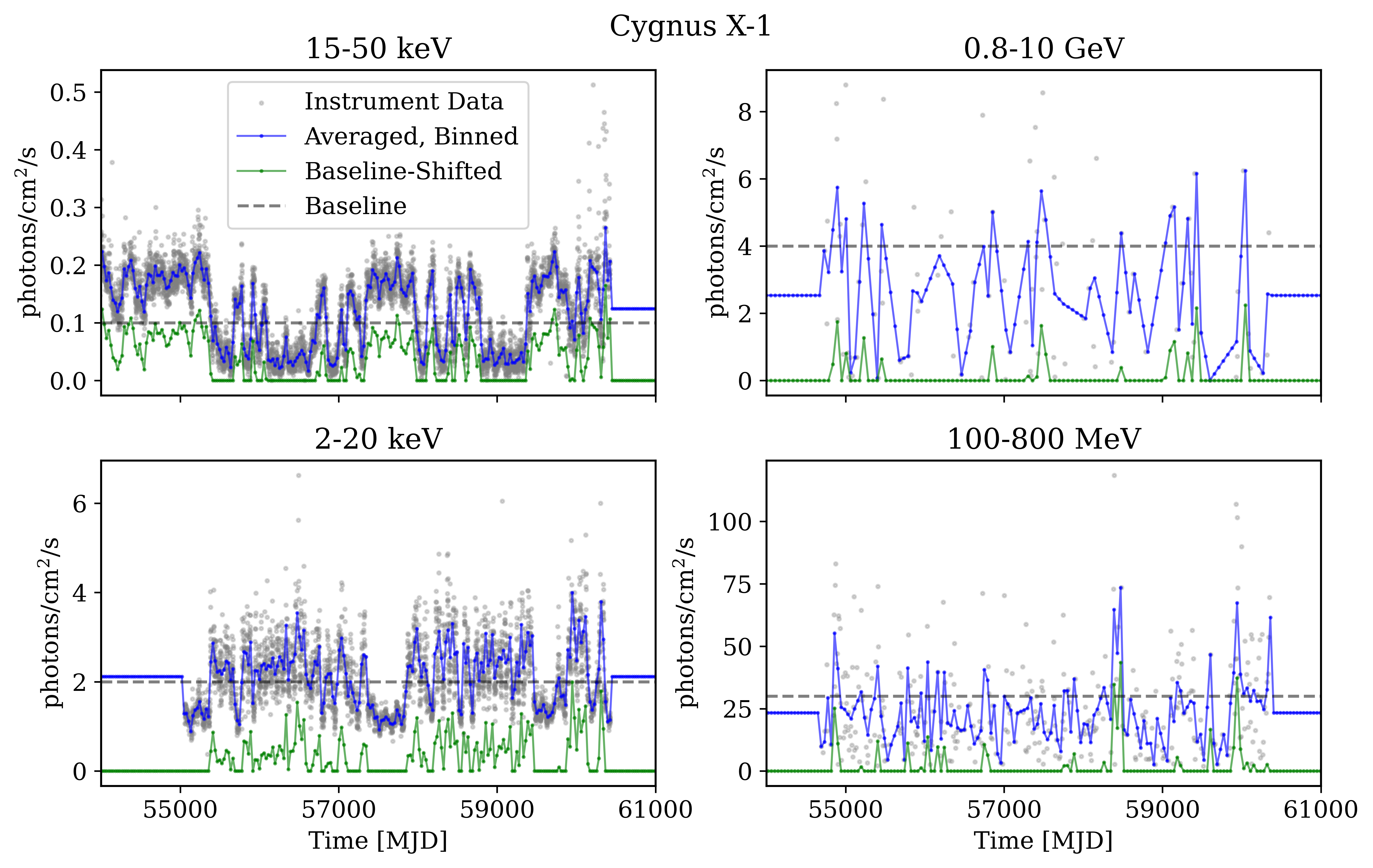}
\caption{Example processed light curves for Cygnus X-1. On the left, example X-ray light curves for hard and soft X-ray bands (top and bottom, respectively) are shown. Similarly, hard and soft gamma-ray observations are pictured at top and bottom right. In all cases, individual observations from the instrument are plotted with gray markers. Uncertainties are not pictured for the clarity of other plotted objects. After these measurements are binned and averaged, the resulting distribution is plotted in blue. A gray, dashed line indicates the threshold flux level, or baseline selected for each source. This level is chosen to differentiate steady-state emission from variable activity. After subtracting the steady-state flux component presumed to be less than this flux level, this resulting baseline-shifted light curve is shown in blue. }\label{fig:lcs}
\end{figure}

All light curves show substantial variability. Both gamma-ray and X-ray emission show high states or flaring periods relative to apparent steady-state emission. As neutrino production may only become efficient during certain spectral states or with jet formation, we also consider correlation without the steady-state flux component. This is accomplished by selecting a `baseline' flux-level, differentiating low-intensity steady-state emission from variable flaring activity. Flux contributions below this threshold are effectively subtracted from the PDF. Examples of processed light curves for Cygnus X-1 prior to normalization are shown in Figure \ref{fig:lcs}. 

In addition to the time-dependent search using flaring light curves, we also perform a periodic analysis for each source. Template light curves are constructed by folding time over the known orbital period of each binary, with an assumed duty cycle of 50$\%$. A time lag of $\pm 1/4$ of the period is allowed. This approach allows us to test for neutrino emission correlated with binary motion and possible periodic modulation of jet activity. 

We perform a likelihood analysis for each source and temporal PDF. The relative, bounded lag between each temporal PDF and neutrino data, $T$, is left free. An astrophysical power-law energy spectrum of free index, $\gamma$, is assumed. These signal parameters are fit along with the number of astrophysical events from the source, $n_s$.    

The unbinned likelihood function incorporates spatial, temporal, and energy information for each detected neutrino event. The construction and application of this likelihood analysis has been motivated in previous work \citep{Braun_2008}. The likelihood function is given by:
\begin{align*}
\mathcal{L}(n_s, \gamma, T) = \prod_{i=1}^{N} \bigg[ \frac{n_s}{N} \cdot S_{\text{temp.}}(t_i|T) \cdot S_{\text{spat.}}(\delta_i, \text{R.A.}_i) \cdot S_{\text{ener.}}(E_i|\delta_i, \gamma)  + \\ \left(1 - \frac{n_s}{N}\right) \cdot \frac{1}{2\pi} \cdot B_{\text{temp.}}(t_i) \cdot B_{\text{dec.}}(\delta_i, t_i) \cdot B_{\text{ener.}}(E_i|\delta_i) \bigg].
\end{align*}
The functions $S_{\text{temp.}}$, $S_{\text{spat.}}$, and $S_{\text{ener.}}$ represent the temporal, spatial and energy signal PDFs, respectively. Similarly, $B_{\text{temp.}}$, $B_{\text{dec.}}$, and $B_{\text{ener.}}$ are the corresponding background distributions. The product runs over all $N$ neutrino candidate events considered in the analysis.

The likelihood test statistic, TS, is defined by the likelihood-maximizing, best-fit signal parameters, $\hat{n}_{s}$, $\hat{\gamma}$ and $\hat{T}$:
\begin{equation}
\textrm{TS} = 2\Bigg[  \dfrac{  \ln\mathcal{L}(n_{s} = \hat{n}_{s}, \gamma = \hat{\gamma}, T = \hat{T})  }{  \ln\mathcal{L}( n_{s} = 0 ) } \Bigg].
\end{equation}

Another strategy for searching for neutrino emission from X-ray binaries is to search for emission from a larger amount of sources stacked together. Under the assumption that all sources emit following a specific weighting, stacking can provide improved sensitivity when compared to individual searches. As an extended part of this analysis, IceCube is also searching for emission from a selection of 75 X-ray binaries that may be weaker emitters. This search will use Swift light curves and sources selected based on variability and excess variance as defined by Swift. These stacked searches weigh the sources by the X-ray flux emitted, and have a similar analysis structure to the previously mentioned search.  
\section{Analysis Performance}\label{sec8}

We determine the IceCube analysis performance for both a correlation with the original, binned temporal models of each source as well as the baseline-subtracted models. Data is randomized as a function of right ascension to generate unique background realizations or trials \citep{Braun_2008}. Background events from atmospheric cosmic rays and unassociated astrophysical sources is expected to be nearly isotropic in right ascension. While an anisotropic diffuse component from the galactic plane is expected, the expected rate of related galactic events is negligible for the purpose of this pointed analysis. Simulated events are added according to an assumed source hypothesis, a specific relative lag, $T$, and spectral index, $\gamma$. 

\begin{figure}
\centering
\begin{subfigure}{.5\textwidth}
  \centering
  \includegraphics[width=\linewidth]{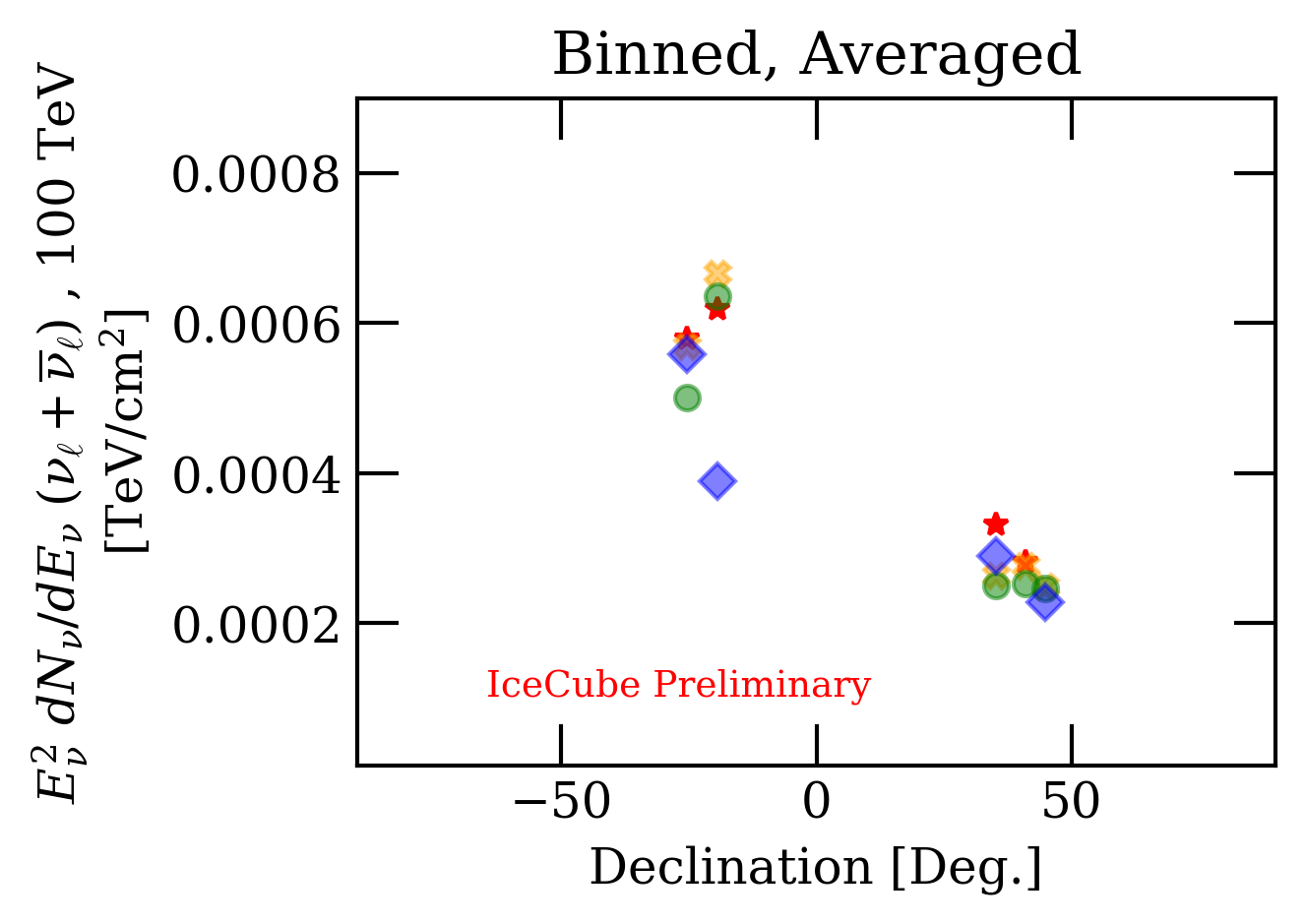}
  \label{fig:sub1}
\end{subfigure}%
\begin{subfigure}{.5\textwidth}
  \centering
  \includegraphics[width=\linewidth]{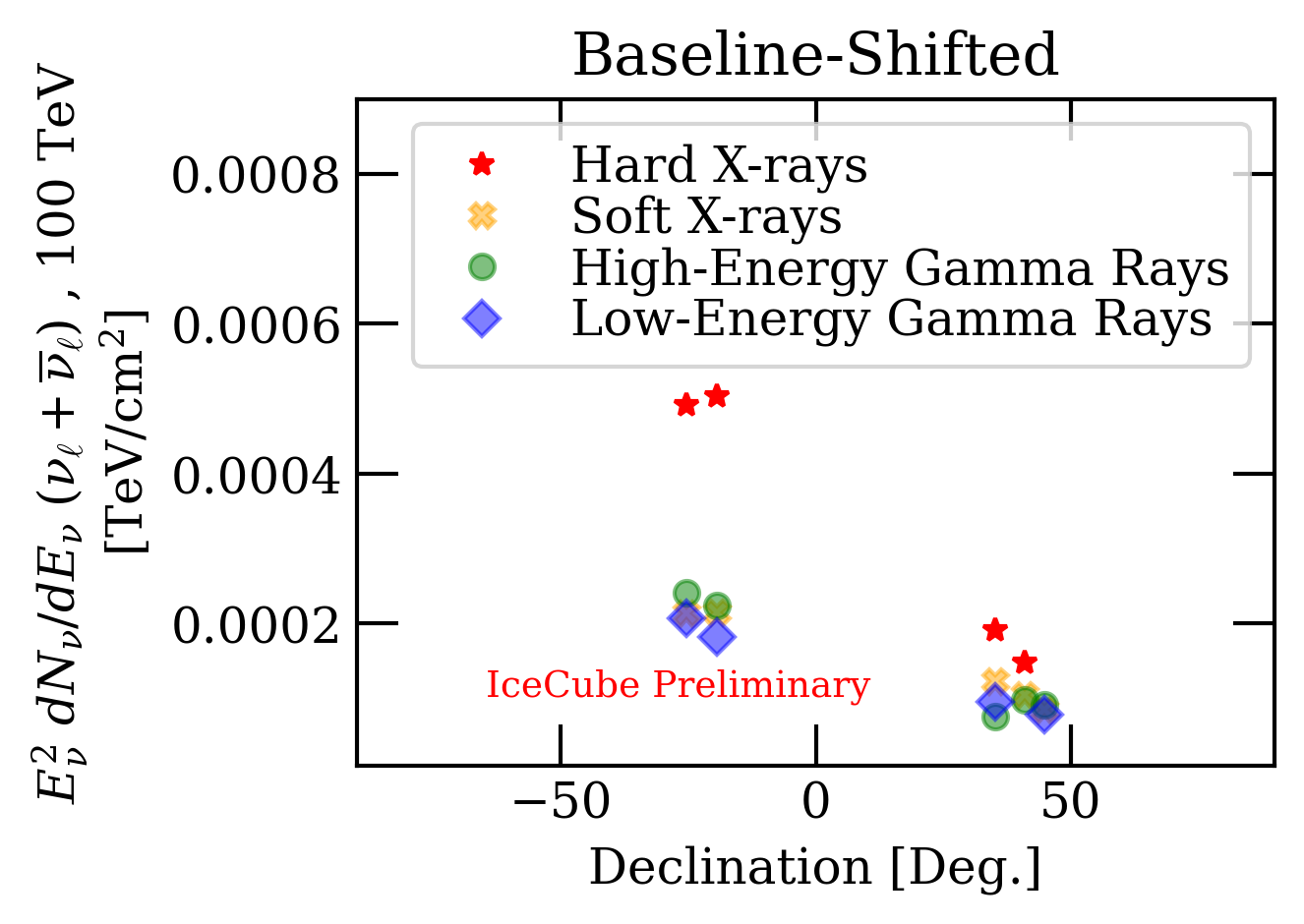}
  \label{fig:sub2}
\end{subfigure}
\caption{3$\sigma$ discovery-potential fluxes for the time-dependent correlations with processed electromagnetic data are pictured. On the left, projections for sources tested with their binned, averaged light curves are shown. On the right, projections for the baseline-shifted temporal model are shown. The IceCube point-source sensitivity in the north is generally superior due to the diminished atmospheric track-like cosmic-ray background. However, the only few-factor difference between northern and southern sources suggests improvement due to the inclusion of cascades.  }
\label{fig:sens}
\end{figure}

An ensemble of background-only data realizations is generated and analyzed. A new set of trials with injected signal of varying intensity, $n_{\textrm{s}}$, is also created. The required neutrino flux at which the test statistic exceeds the median of the background test-statistic distribution in 90$\%$ of the cases corresponds to the sensitivity flux. The 3$\sigma$ or 5$\sigma$ discovery potential indicates the flux level at which in 50$\%$ of the cases the test statistic exceeds three or five standard deviations of the background test statistic distribution, respectively. 3$\sigma$ discovery-potential fluxes for the averaged, binned and baseline-shifted light-curve correlations are shown in Figure \ref{fig:sens}. Projections for the periodic temporal test are shown in Table \ref{tab:period}.

We also note that under the previously mentioned X-ray flux-weighted stacked analysis, a five-sigma discovery potential of 3.7 $\times$ $10^{-4}$ $\mathrm{TeV}$ $\mathrm{cm}^{-2}$ at 100 TeV is found for an assumed power-law energy spectrum of index 2.

\begin{table}[]
\centering
\label{tab:sensitivity_table}
\begin{tabular}{ccccc}
\hline
Source Name & Sensitivity ($\gamma = 2$) & D.P. ($\gamma = 2$) & Sensitivity ($\gamma = 3$) & D.P. ($\gamma = 3$) \\
\hline
Cygnus X-1 & 1.25 & 4.41 & 0.217 & 1.10 \\
Cygnus X-3 & 1.40 & 5.17 & 0.24 & 0.97 \\
V4641 Sagittarii & 3.80 & 13.53 & 3.00 & 9.81 \\
MAXI J1836-194 & 3.07 & 12.59 & 3.23 & 10.49 \\
MWC 656 & 1.29 & 4.81 & 0.21 & 1.17 \\
\hline
\end{tabular}
\caption{Periodic temporal-model sensitivities for the selected sources of this work. Sensitivity and 5$\sigma$ discovery-potential (D.P.) fluxes are reported for a simulated power-law energy spectrum either of index, $\gamma$ = 2 or 3. Corresponding energy fluxes, $\phi$, are presented in the form $E^{2}_{\nu}$ $dN_{\nu}/dE_{\nu}(\mathrm{100 \,TeV}) = \phi \, \times$ $10^{-4}$\,$\mathrm{TeV}$ $\mathrm{cm}^{-2}$.}
\label{tab:period}
\end{table}

\section{Conclusion}\label{sec9}
We have considered a selection of galactic X-ray binaries based on their feasibility as observable neutrino sources. We have developed a set of time-dependent likelihood analyses utilizing hard and soft Fermi-LAT gamma-ray observations as well as MAXI and Swift X-ray data. The analysis will probe potential neutrino emission correlated both with jet production and with varying accretion states. The analysis may have the ability to first probe or even tease apart potential neutrino contributions associated with both the jet and coronal region around the disk. The results of this search will follow in a future publication. 

\bibliographystyle{ICRC}
\bibliography{references}

\clearpage

\section*{Full Author List: IceCube Collaboration}

\scriptsize
\noindent
R. Abbasi$^{16}$,
M. Ackermann$^{63}$,
J. Adams$^{17}$,
S. K. Agarwalla$^{39,\: {\rm a}}$,
J. A. Aguilar$^{10}$,
M. Ahlers$^{21}$,
J.M. Alameddine$^{22}$,
S. Ali$^{35}$,
N. M. Amin$^{43}$,
K. Andeen$^{41}$
% ...
\\
\\
% ...
$^{10}$ Universit{\'e} Libre de Bruxelles, Science Faculty CP230, B-1050 Brussels, Belgium \\
% ...
$^{16}$ Department of Physics, Loyola University Chicago, Chicago, IL 60660, USA \\
$^{17}$ Dept. of Physics and Astronomy, University of Canterbury, Private Bag 4800, Christchurch, New Zealand \\
% ...
$^{21}$ Niels Bohr Institute, University of Copenhagen, DK-2100 Copenhagen, Denmark \\
$^{22}$ Dept. of Physics, TU Dortmund University, D-44221 Dortmund, Germany \\
% ...
$^{35}$ Dept. of Physics and Astronomy, University of Kansas, Lawrence, KS 66045, USA \\
% ...
$^{39}$ Dept. of Physics and Wisconsin IceCube Particle Astrophysics Center, University of Wisconsin{\textemdash}Madison, Madison, WI 53706, USA \\
% ...
$^{41}$ Department of Physics, Marquette University, Milwaukee, WI 53201, USA \\
% ...
$^{43}$ Bartol Research Institute and Dept. of Physics and Astronomy, University of Delaware, Newark, DE 19716, USA \\
% ...
$^{63}$ Deutsches Elektronen-Synchrotron DESY, Platanenallee 6, D-15738 Zeuthen, Germany \\
$^{\rm a}$ also at Institute of Physics, Sachivalaya Marg, Sainik School Post, Bhubaneswar 751005, India \\
% ...

\subsection*{Acknowledgments}

\noindent
The authors gratefully acknowledge the support from the following agencies and institutions:
USA {\textendash} U.S. National Science Foundation-Office of Polar Programs,
U.S. National Science Foundation-Physics Division,
U.S. National Science Foundation-EPSCoR,
U.S. National Science Foundation-Office of Advanced Cyberinfrastructure,
Wisconsin Alumni Research Foundation,
Center for High Throughput Computing (CHTC) at the University of Wisconsin{\textendash}Madison,
Open Science Grid (OSG),
Partnership to Advance Throughput Computing (PATh),
Advanced Cyberinfrastructure Coordination Ecosystem: Services {\&} Support (ACCESS),
Frontera and Ranch computing project at the Texas Advanced Computing Center,
U.S. Department of Energy-National Energy Research Scientific Computing Center,
Particle astrophysics research computing center at the University of Maryland,
Institute for Cyber-Enabled Research at Michigan State University,
Astroparticle physics computational facility at Marquette University,
NVIDIA Corporation,
and Google Cloud Platform;
Belgium {\textendash} Funds for Scientific Research (FRS-FNRS and FWO),
FWO Odysseus and Big Science programmes,
and Belgian Federal Science Policy Office (Belspo);
Germany {\textendash} Bundesministerium f{\"u}r Forschung, Technologie und Raumfahrt (BMFTR),
Deutsche Forschungsgemeinschaft (DFG),
Helmholtz Alliance for Astroparticle Physics (HAP),
Initiative and Networking Fund of the Helmholtz Association,
Deutsches Elektronen Synchrotron (DESY),
and High Performance Computing cluster of the RWTH Aachen;
Sweden {\textendash} Swedish Research Council,
Swedish Polar Research Secretariat,
Swedish National Infrastructure for Computing (SNIC),
and Knut and Alice Wallenberg Foundation;
European Union {\textendash} EGI Advanced Computing for research;
Australia {\textendash} Australian Research Council;
Canada {\textendash} Natural Sciences and Engineering Research Council of Canada,
Calcul Qu{\'e}bec, Compute Ontario, Canada Foundation for Innovation, WestGrid, and Digital Research Alliance of Canada;
Denmark {\textendash} Villum Fonden, Carlsberg Foundation, and European Commission;
New Zealand {\textendash} Marsden Fund;
Japan {\textendash} Japan Society for Promotion of Science (JSPS)
and Institute for Global Prominent Research (IGPR) of Chiba University;
Korea {\textendash} National Research Foundation of Korea (NRF);
Switzerland {\textendash} Swiss National Science Foundation (SNSF).

\end{document}